\documentclass[10pt, conference, compsocconf]{IEEEtran}

\usepackage[utf8]{inputenc} % allow utf-8 input
\usepackage[T1]{fontenc}    % use 8-bit T1 fonts
\usepackage{url}            % simple URL typesetting
\usepackage{booktabs}       % professional-quality tables
\usepackage{amsfonts}       % blackboard math symbols
\usepackage{nicefrac}       % compact symbols for 1/2, etc.
\usepackage{microtype}      % microtypography
\usepackage{lipsum}
\usepackage{graphicx}
\usepackage{subcaption}
\usepackage{amsmath}
\usepackage{tikz}
\usepackage{listings}
\usepackage{dblfloatfix}
\usetikzlibrary{automata}
\graphicspath{ {./images/} }

\makeatletter
\def\BState{\State\hskip-\ALG@thistlm}
\makeatother

\begin{document}
\title{Automated Primary Hyperparathyroidism Screening with Neural Networks}

\author{\IEEEauthorblockN{Noah Ziems}
\IEEEauthorblockA{Department of Computer Science \\
Ball State University\\
Muncie, IN USA\\
nmziems@bsu.edu}
\and
\IEEEauthorblockN{Shaoen Wu}
\IEEEauthorblockA{School of Information Technology \\
Illinois State University\\
Normal, IL USA \\
swu1235@ilstu.edu}
\and
\IEEEauthorblockN{Jim Norman}
\IEEEauthorblockA{
Norman Parathyroid Center\\
Tampa, FL USA \\
jimnormanmd@gmail.com}
}

\maketitle

\begin{abstract}

Primary Hyperparathyroidism(PHPT) is a relatively common disease, affecting about one in every 1,000 adults. However, screening for PHPT can be difficult, meaning it often goes undiagnosed for long periods of time. While looking at specific blood test results independently can help indicate whether a patient has PHPT, often these blood result levels can all be within their respective normal ranges despite the patient having PHPT. Based on the clinic data from the real world, in this work, we propose a novel approach to screening PHPT with neural network (NN) architecture, achieving over 97\% accuracy with common blood values as inputs. Further, we propose a second model achieving over 99\% accuracy with additional lab test values as inputs. Moreover, compared to traditional PHPT screening methods, our NN models can reduce the false negatives of traditional screening methods by 99\%.

\end{abstract}

\section{Introduction} \label{intro}
Primary Hyperparathyroidism (PHPT) is a widely existing disease caused when a parathyroid develops a tumor causing the parathyroid to over produce parathyroid hormone (PTH), which regulates the amount of calcium in the bloodstream. Too much calcium in the bloodstream for extended periods of time leads to various health issues such as osteoporosis, kidney dysfunction, high blood pressure, and others. Moreover, PHPT is relatively common, affecting about one in every 1,000 adults. However, the current medical practice of manual screening for PHPT is very difficult. While identifying whether a patient has PHPT can be facilitated by independently inspecting a few variables of their blood lab results, often these blood test variable values can all be within normal ranges despite a patient having PHPT. A more reliable way to screen for PHPT, therefore, is to consider the values of the blood test variables coherently as opposed to independently. Because these blood variable values in lab results are not often significant individually, millions of patients with PHPT have been missed in their traditional annual physical exams that have blood lab results. Only when they show PHPT symptoms later will their lab results will be examined by a specialist that can diagnose PHPT. Therefore, it is of utmost importance to design an automated AI solution to diagnose the patients with PHPT from their blood lab results in regular visits or physical exams much earlier than symptoms begin to show. 

Deep learning with neural networks(NNs) has been shown to be surprisingly effective at learning relationships among data, often surpassing human performance in data-driven tasks such as object detection and recognition ~\cite{russakovsky2015imagenet}. Deep learning driven health informatics in many areas such as cancer diagnosis has achieved amazing progress in recent years with many solutions reportedly outperforming human medical experts ~\cite{esteva2021}. In this work, we present a novel machine learning based solution for screening PHPT in patients using simple neural networks. This work is a collaboration between computer science and a world-class PHPT clinic with the real medical lab result data of thousands of patients.

More specifically, our solution demonstrates the following advantages over traditional PHPT screening methods:
\begin{enumerate}
    \item \textbf{Speed}: By using automated algorithmic methods that can be run on a computer, each individual patient is evaluated in a matter of milliseconds where a trained doctor may take minutes or more to interpret in traditional screening methods.
    \item \textbf{Cost}: The computational cost of evaluating each patient is significantly less than \$0.01, while using a doctor with traditional screening methods costs significantly more.
    \item \textbf{Scale}: Instead of evaluating patients one by one as a doctor would, our model is able to run in parallel, evaluating many patients at once.
    \item \textbf{Accuracy}: Our method proves itself to be highly accurate while allowing for future improvements with more data. However, it remains to be seen how this accuracy compares to that of an experienced medical expert.
    \item \textbf{Precision}: Due to the deterministic nature of computer software, the same results are always given for the same inputs. Where doctors may differ in opinion, our method will always produce the same diagnosis given the same inputs.
\end{enumerate}

\section{Related Work}\label{sec:related}

\subsection{Machine Learning in Health Informatics}\label{subsec:ml-healthcare}
Significant progress has been made using state-of-the-art deep learning based computer vision techniques in health informatics \cite{bellamy2020evaluating}. Wu et al. develop a computer vision model, finding it as accurate as experienced radiologists in screening breast cancer \cite{wu2019deep}. Moreover, there are a number of benchmark datasets that have shown impressive improvements in performance for disease diagnosis, including CheXpert \cite{irvin2019chexpert}, SD-198\cite{sun2016benchmark}, and the International Skin Imaging Collaboration (ISIC) dataset \cite{rotemberg2020patient}.

However, relatively little progress has been made on more traditional tabular data that most medical data is comprised of, typically in the form of Electronic Health Records (EHRs). There is some debate as to why machine learning has not had as much impact in health informatics as in other fields. New studies show this is likely caused by a lack of benchmarks in the field, preventing machine learning researchers from comparing new methods to traditional ones \cite{bellamy2020evaluating}.

\subsection{PHPT Screening in Endocrinology}\label{subsec:screen-endocrine}
If there is no prior suspicion of PHPT, screening of PHPT is only done when the patient has a metabolic panel drawn to check his or her general health. A high serum calcium level from a metabolic panel is often an indication of PHPT. However, very often doctors disregard this because many other factors can contribute to high calcium levels. The true normal range of serum calcium varies based on other factors, such as age and gender, which are not often taken into account by traditional screening methods. Instead, these traditional screening methods only evaluate a patient based on the normal range of a blood lab value for the general population irrespective of the other blood lab values. How to accurately diagnose PHPT in an integrative consideration of all these factors remains open, which also motivates our work of achieving such a goal with machine learning.

\section{Clinical Dataset}\label{sec:data}

In this work, we propose using machine learning to accurately diagnose PHPT by considering the factors including serum calcium level that can be obtained in blood lab results, age and gender. The data supporting our machine learning model consists of two separate parts. The first part comes from real PHPT patient data while the second part is synthetic data generated using the observed distribution metrics of patients without PHPT. 

The data stemming from real patients comes from the Norman Parathyroid Center \cite{norman2011}, which annually conducts the most parathyroid surgeries in the United States. The dataset is comprised of the data from 20,000 patients having positive diagnosis of PHPT. In the data, each record consists of parameters including age, gender, pre-operative serum calcium level, and pre-operative parathyroid (PTH) hormone, all of which are the input to our machine learning PHPT screening model, as well as some other information about each patient. The extra information is not taken into account by our model because their measurements are only taken after a patient is suspected to have PHPT and thus would not fall into the category of screening. It is worth noting that each patient in the dataset had the serum calcium and parathyroid hormone levels measured at least times which are then averaged before they are put into the dataset. During training time, the record of each patient in this portion of the dataset is labeled as having PHPT. In the dataset, patient privacy information such as name and address has been stripped off.

To supplement our dataset with non-PHPT data, which is necessary for training a machine learning model, we have generated synthetic data based on the previously learned distributions of serum calcium and parathyroid hormone in patients without PHPT. Each synthetic data point is given a calcium level drawn with a Gaussian distribution of $\mu=9.6$mg/dl and $\sigma=0.15$ and a PTH level with a Gaussian distribution of $\mu=34$pg/ml and $\sigma=4.5$ \cite{norman2011}. To maintain a dataset that is independent and identically distributed(IID), the synthetic data has the exactly same number of data points as the real patient data, yielding a total dataset of around 40,000 data points.

Based on the distributions our synthetic data is drawn from, half of the data labeled as PHPT-Negative is male with the other half being female. For those who are labeled in the dataset as PHPT-Positive, 76\% are female. This may appear to indicate a dataset which is not independently and identically distributed (IID). However, this proportion is in line with previously observed distribution of PHPT Positive incidence in clinical settings \cite{yeh2013}. 

\begin{figure*}[!hbt]
    \centering
    \begin{subfigure}[b]{0.3\textwidth}
      \includegraphics[width=\textwidth]{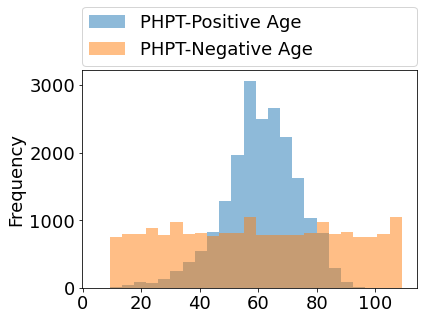}
      \centering
      \caption{The distributions of age in our dataset. Note the PHPT-Positive distribution is from real patients with PHPT where the PHPT-Negative distribution is generated from a uniform distribution.}
      \label{fig:age-histograms}
    \end{subfigure}
    \hspace{0.1in}
    \begin{subfigure}[b]{0.3\textwidth}
      \includegraphics[width=\textwidth]{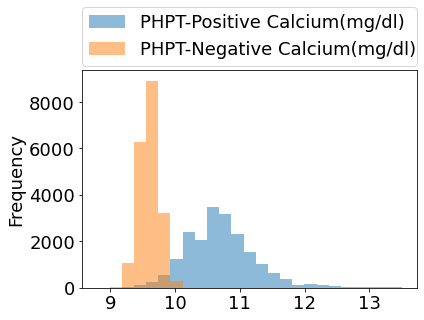}
      \caption{The distributions of calcium in our dataset. Note the PHPT-Positive distribution is from real patients with PHPT where the PHPT-Negative distribution is generated from a normal distribution.}
      \label{fig:calcium-histograms}
    \end{subfigure}
    \hspace{0.1in}
    \begin{subfigure}[b]{0.3\textwidth}
      \includegraphics[width=\textwidth]{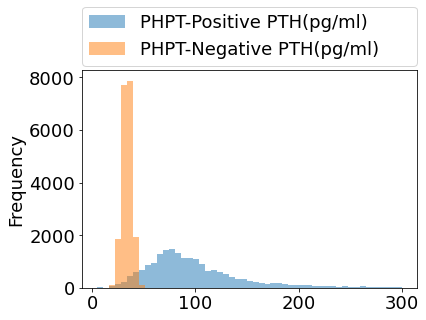}
      \caption{The distributions of PTH in our dataset. Note the PHPT-Positive distribution is from real patients with PHPT where the PHPT-Negative distribution is generated from a normal distribution.}
      \label{fig:pth-histograms}
    \end{subfigure}
    \caption{The distributions of age, calcium and PTH in the clinic dataset}
      \label{fig:distributions}
\end{figure*}

Figure~\ref{fig:distributions} shows the distributions of the important continuous data used by our models. Areas in blue indicate PHPT-Positive distributions where areas in orange indicate PHPT-Negative distributions. The distributions have relatively little overlap, which further indicates machine learning is a promising solution for PHPT screening. It is worth noting that the distribution of PHPT-Negative in Figure~\ref{fig:age-histograms} is intentionally uniform from the assumption that age of the general population is uniform as opposed to Gaussian.

\section{Neural Network Screening for Primary Hyperparathyroidism}\label{sec:architecture}

\begin{figure}[!ht]
\centering
  \includegraphics[width=0.35\textwidth]{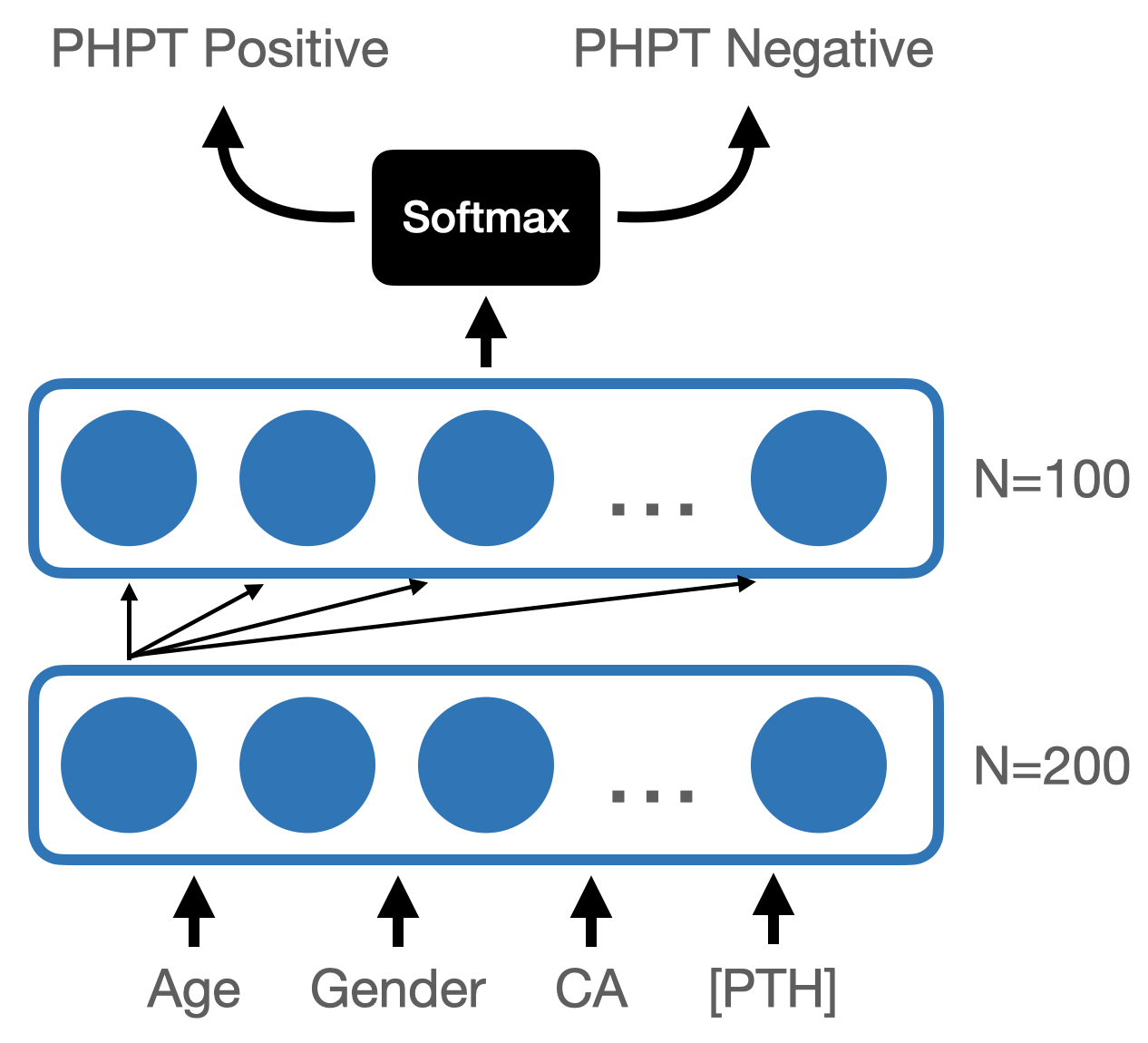}
  \caption{The generic neural network architecture used by both models}
  \label{fig:phpt-positive-gender}
\end{figure}

This paper proposes using neural network (NN) models to automate the PHPT screening based on four factors: age, gender, pre-operative serum calcium level, and pre-operative parathyroid (PTH) hormone level. Due to the relatively low dimensionality of the input data and output data, a simple neural network architecture has been chosen. We have designed and implemented two nearly-identical NN models for PHPT screening, with the only difference being the number of inputs. The generic architecture of the PHPT NN is illustrated on Figure~\ref{fig:phpt-positive-gender}.

\subsection{PHPT Neural Network Inputs}\label{subsec:inputs}
For both models, gender, age, and average serum calcium(Avg Ca) are taken as inputs. However, one of the models also takes average parathyroid hormone(PTH) into account while the other does not. To avoid any confusion, we refer to these models as \emph{PTH-Included} and \emph{PTH-Ignored}. For both models, gender is treated as a categorical variable that is one-hot encoded. Age, average calcium, and average PTH are all treated as continuous inputs.

The motivation of designing two different identical NN models lies in medical application use cases. The goal of this work is to improve screening for PHPT. PTH is a critical component in diagnosing PHPT, however it is rarely tested if there is no prior suspicion of PHPT. In contrast, serum calcium is far more commonly tested in regular blood lab work. Therefore, we create another model, \emph{PTH-Ignored}, which only takes inputs that are common in regular lab tests for patients who are not suspected to have PHPT. The \emph{PTH-Ignored} model can then be used to screen a much broader set of patients in hopes of finding more PHPT patients even if there is no prior suspicion of PHPT.

\subsection{PHPT Neural Network Hidden Layers}\label{subsec:hidden}
As stated above, both NN model architectures are identical in their internal structures. They both have two hidden layers with the first layer containing 200 neurons and the second layer containing 100 neurons. A ReLU\cite{nairhinton2010} is used in between these two layers, serving as the necessary nonlinear activation function.

\subsection{PHPT Neural Network Outputs}\label{subsec:outputs}
The output of the last hidden layer is a vector that is run through a Softmax function to find the final classification probabilities as the final output of the NN for PHPT screening. For both models, there are only two output classes, one for PHPT-Positive and the other for PHPT-Negative.

\subsection{PHPT Neural Network Model Training}\label{sec:training}
For the PHPT NN model training, the dataset is split using standard 5-fold cross validation. The training set accounts for 80\% of the original dataset and the test set accounts for the remaining 20\%. For each training iteration, the patient lab values in the training dataset are given as input to the models. After the models computes the outputs, a loss function is used to measure the deviation of the prediction from its real value. The loss then backpropagates through the models and weights are updated so the models are more correct next time they are given similar data as inputs. After many thousands of iterations, the models converge on a particular set of weights that has minimal loss. In each of our experiments, the models are trained for five epochs with a triangular cyclical learning rate and minimal hyperparameter tuning. It is worth noting that both models are trained in the exactly same way with the only difference being on the input of the PTH values or not.

\section{Performance Evaluation}\label{sec:results}

\begin{figure*}[htb]
    \centering
    \begin{subfigure}[b]{0.4\textwidth}
      \includegraphics[width=0.8\textwidth]{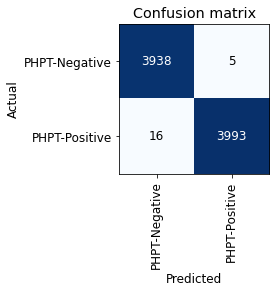}
      \caption{A confusion matrix showing our model's predictions versus the ground truth of the PHT-Included model}
      \label{fig:phpt-ca-pth-confusion-matrix}
    \end{subfigure}
    \hspace{1cm}
    \begin{subfigure}[b]{0.4\textwidth}
      \includegraphics[width=0.8\textwidth]{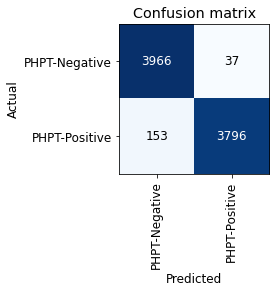}
      \caption{A confusion matrix showing our model's predictions versus the ground truth of the PHT-Ignored model}
      \label{fig:phpt-ca-only-confusion-matrix}
    \end{subfigure}
        \caption{Confusion Matrices}
    \label{fig:confusion}
\end{figure*}

\subsection{Experiment Platform and Settings}
We have implemented and evaluated our PHPT NN models with PyTorch \cite{NEURIPS2019_9015}. On the top of PyTorch we use FastAI \cite{howard2018fastai} due to it's existing API designed for tabular data, which evaluating our dataset requires. After training the models on the training dataset, we then assessed them on the test dataset, which is data the trained models have never seen before. This test dataset is meant to represent the data the models would come across when used in practice. 

For fair comparison, we have evaluated the same dataset using traditional screening methods. For traditional PHPT screening, any calcium level between 8.4 mg/dl and 10.5 mg/dl is considered normal and anything outside of that is flagged as abnormal. For PTH, anything between 9 pg/ml and 69 pg/ml is considered normal while anything outside that range is considered abnormal.

\subsection{Performance Metrics}

The key performance metrics used to evaluate our models are \textit{accuracy, precision}, and \textit{recall}. \textit{Accuracy} in our case indicates how often our models' prediction is correct in recognizing a patient has PHPT based on the input clinic data. Higher accuracy is better. \textit{Precision} indicates the percentage of positive predictions our models correctly predict a patient has PHPT based the input data. \textit{Precision} is much more sensitive to false positives than accuracy. \textit{Recall} is very similar to \textit{Precision}, but is instead sensitive to false negatives. The definition and calculation for \textit{accuracy}, \textit{precision}, and \textit{recall} are shown below.

$$Accuracy = \frac{TP+TN}{TP+TN+FP+FN}$$
$$Precision = \frac{TP}{TP+FP}$$
$$Recall = \frac{TP}{TP+FN}$$

Here $TP$ stands for true positive, $TN$ for true negative, $FP$ for false positive and $FN$ for false negative. 

\subsection{Confusion Matrices}
The confusion matrices in Figure~\ref{fig:confusion} show all data needed to compute the results shown in Table \ref{tab:results}. The numbers in the top left and bottom right in dark blue are $TP$ and $TN$ indicating how many datapoints are correctly classified in the dataset whereas the numbers in the top right and bottom left in light grey are $FP$ and $FN$ indicating how many datapoints in the dataset are misclassified. As shown in Figure \ref{fig:phpt-ca-pth-confusion-matrix}, the PTH-Included model has significantly fewer false negatives when compared to the PTH-Ignored model shown in Figure \ref{fig:phpt-ca-only-confusion-matrix}.

\subsection{Results and Observations}
The results of our experiments are summarized in Table~\ref{tab:results}.

\begin{table}[!ht]
    \centering
    \begin{tabular}{lllll}
    \hline
    Architecture  & Accuracy & Precision & Recall    \\
    \hline
    PTH-Included  & 99.73\%  & 99.87\%   & 99.60\%   \\
    \hline
    PTH-Ignored   & 97.61\%  & 99.03\%   & 96.12\%   \\
    \hline
    Traditional Screening   & 93.06\%  & 100\%   & 88.69\%   \\
    \hline
    \end{tabular}
    \caption{Performance summary of PHPT screening with two NN models and traditional screening}
    \label{tab:results}
\end{table}

From the results, it can be observed that the screening accuracy of both models far surpassed the performance of traditional screening methods. It is worth noting that the PTH-Included model performed significantly better than the PTH-Ignored model, as the PTH-Included model has access to more information relevant to diagnosis than the PTH-Ignored model. This is an intentional design choice.

A surprising observation is that, although the traditional screening method had 0 false positives after evaluating all patients in the dataset, it had over 2,000 false negatives. In other words, a significant portion of patients with PHPT in the dataset would not have been recognized as having PHPT. In contrast, the PHPT-Included NN model only had 5 false positives and 16 false negatives. Therefore, our best PHPT NN model can reduce the false negatives of traditional expert-screening by 99.27\%.

The accuracy of both PHPT NN models suggest that screening for PHPT with machine learning is not only possible, but could be incredibly effective in practice. Moreover, the total marginal cost of evaluating a single patient is only from the cost of electricity to support the computation of the model that takes on average seven milliseconds on our current hardware.

\section{Model Capabilities and Limitations}\label{sec:limitation}

\begin{figure*}
    \centering
    \begin{subfigure}[b]{0.3\textwidth}
      \includegraphics[width=\textwidth]{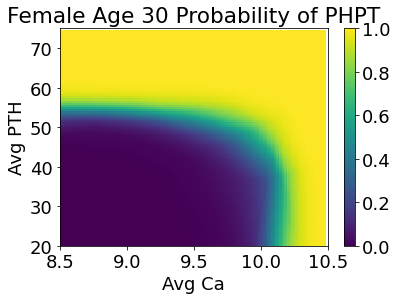}
      \caption{A prediction map of PHPT given a female age 30. Yellow indicates high probability of PHPT while purple indicates low probability.}
      \label{fig:phpt-map-female-age-30}
    \end{subfigure}
    \hspace{0.1cm}
    \begin{subfigure}[b]{0.3\textwidth}
      \includegraphics[width=\textwidth]{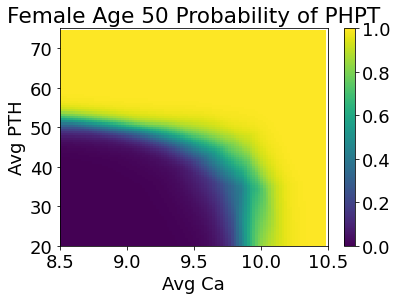}
      \caption{A prediction map of PHPT given a female age 50. Yellow indicates high probability of PHPT while purple indicates low probability.}
      \label{fig:phpt-map-female-age-50}
    \end{subfigure}
\hspace{0.1cm}
    \begin{subfigure}[b]{0.3\textwidth}
      \includegraphics[width=1.15\textwidth]{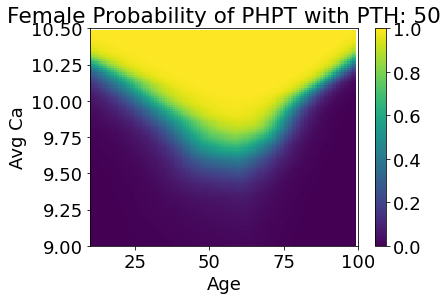}
      \caption{A prediction map of PHPT given a female with PTH 50. Yellow indicates high probability of PHPT while purple indicates low probability.}
      \label{fig:phpt-map-female-pth-50}
    \end{subfigure}
\hspace{1cm}
    \begin{subfigure}[b]{0.35\textwidth}
      \includegraphics[width=\textwidth]{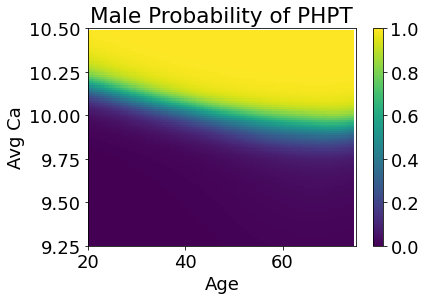}
      \caption{A prediction map of PHPT for a male. Note this is for the PTH-Ignored model.}
      \label{fig:phpt-map-ca-only-male}
    \end{subfigure}
    \hspace{1cm}
    \begin{subfigure}[b]{0.35\textwidth}
      \includegraphics[width=\textwidth]{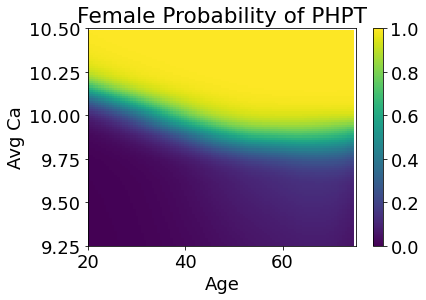}
      \caption{A prediction map of PHPT for a female. Note this is for the PTH-Ignored model.}
      \label{fig:phpt-map-ca-only-female}
    \end{subfigure}
    \caption{Model Predication Maps}
    \label{tab:maps}
\end{figure*}

The diagrams on Figure~\ref{tab:maps} are prediction maps showing how the model's predictions change when given different inputs. These can also be thought of as the decision boundaries, where yellow indicates high confidence that a patient has PHPT and blue indicates high confidence that a patient does not have PHPT. Green areas indicate where the model is unsure whether the patient has PHPT or not based on the data given.

\subsection{Model Capabilities}
Shown in Figure \ref{fig:phpt-map-female-age-30} and  Figure \ref{fig:phpt-map-female-age-50}, our model begins flagging patients as PHPT positive when calcium goes above 10.5 mg/dl or when PTH goes above 55pg/ml. However, the decision boundary changes shape as the input age increases, flagging patients as PHPT Positive with lower calcium and PTH levels. This is inline with the observed PHPT incidence rate as age increases, particularly with post-menopausal women\cite{yeh2013}. Traditional screening methods do not often take this into account, ignoring age as a factor all together.

Moreover, Figure \ref{fig:phpt-map-ca-only-male} and Figure \ref{fig:phpt-map-ca-only-female} show the PTH-Ignored model's prediction map when comparing male and female patients. The decision boundary is slightly lower with females than males, which is also in line with observed distributions in clinical data. In both cases, the CA level needed to flag a patient as PHTP positive decreases with age, with the decrease being slightly more aggressive for females around age 40.

It is worth noting all of these decision boundaries are statistically learned with no input from the authors aside from the generated synthetic data.

\subsection{Model Limitations}
We have further investigated the limitations and potential issues of the PHPT NN models when the training data is out of expected distribution. As shown in Figure \ref{fig:phpt-map-female-pth-50}, the models can yield mistakes when exposed to data that is substantially out of the distribution it has been trained on. For age 60 and older, the probability of PHPT begins to decrease even when average serum calcium levels increase. To our best knowledge, this is medically incorrect and likely caused by the uniform distribution that was used for age in the synthetic portion of our dataset. Further, when the average serum calcium goes below 9.0 as in Figure \ref{fig:phpt-map-ca-only-male}, the model begins increasingly classifying patients as PHPT-Positive. This also is likely incorrect. Both of these problems would be solved with real data sourced from patients that do not have PHPT, rather than using the synthetic data.

\section{Conclusions}\label{sec:conclusion}
This paper proposes an automated PHPT screening solution with neural network machine learning models. With the real clinic data, the solution shows the ability of simple neural networks to achieve surprising when trained to screen for PHPT. Compared to traditional screening, our solution can significantly improve the accuracy while it can reduce the false negative by 99\%. Moreover, we show our approach to screening for PHPT has numerous advantages to the traditional methods, including speed, scale, and consistency. The robustness and reliability of the model can be even more improved with real data of patients without PHPT. 

\section*{Acknowledgments}
This material is based upon work supported by the National Science Foundation under Grant No. \#2109971.

\bibliographystyle{abbrv}
\bibliography{reference.bib}
\end{document}